\documentclass{article}
\usepackage[utf8]{inputenc}
\usepackage{braket}
\usepackage{authblk}
\usepackage{hyperref}
\usepackage{bm}

\title {Review of Berger and DiRuggiero, “Einstein: The Man and His Mind”}

\author{Galina Weinstein}
\affil{\normalsize Reichman University, The Efi Arazi School of Computer Science, Herzliya; University of Haifa, The Department of Philosophy, Haifa, Israel.} 

\begin{document}

\maketitle

\emph{Einstein: The Man and His Mind} is a large-format, visual coffee-table book by physician and Albert Einstein collector Gary S. Berger and bookseller Michael DiRuggiero. 

The book consists of reproductions of original photographs of Einstein, mostly signed portrait photos, supplemented by holograph letters—letters written entirely in the handwriting of Einstein—and typewritten letters signed by Einstein. The book also contains photographs of original journals, final versions of printed papers, and other documents written in Einstein’s hand. All these original documents were selected from Berger’s private collection in Chapel Hill, North Carolina (probably the largest private collection of Einstein-related documents), and from what DiRuggiero calls the “Einstein Archives,” rare Einstein items for sale in his bookstore, Manhattan Rare Book Company, in New York City.

The photos and items in the book are reproduced in the highest quality. When you look at the book you really think that the photographs, letters, and handwritten documents in the book are originals. The book is printed on thick matte paper and is visually attractive; it looks like an impressive art catalog. The book has special artistic features: the authors put tracing pages on top of several pages containing photos of Einstein. This feels reminiscent of a photo album. There are large-format equations in Einstein’s hand and quotes by Einstein on these tracing pages. Further, two double pages open out from the middle to reveal breathtaking photos of journals and of Einstein standing near a blackboard and pondering his equations.

These photos show Einstein standing near the blackboard writing equations, and they are entitled "Einstein at Work".\footnote{The photos show Einstein working on a generalization of Theodor Kaluza’s theory, adding extra $\Gamma$s to his field equations. $\Gamma$ represents the Christoffel symbols of the second kind written in terms of the metric tensor $g_{mn}$.} They were photographed in 1942 at Princeton by the Russian American photographer Roman Vishniac.
Berger and DiRuggiero talk about the photographic qualities of the photos. One can find these photos in DiRuggiero’s “Einstein Archives” website, and an added description says that in a handwritten letter Vishniac writes: “All pictures are made with ‘hidden camera’ method”. Berger and DiRuggiero add anecdotal remarks pertaining to Vishniac and his photographs.

As in other coffee-table art books, the authors limit the amount of text: there is an item—a photograph, letter, or original document—on one page and a short text discussing it in the context of Einstein’s life, science, philosophy, and political views on the other page. For instance, the first page after the foreword contains a small photo of Einstein in Aarau (the capital of the Swiss canton of Aargau) from DiRuggiero’s “Einstein Archives” on the left-hand page. The photo is relatively small because it is true to the original size (6.5 by 10.5 centimeters). A short text on the right-hand page says, “This is the earliest known signed photograph of Albert Einstein.” It was taken when he was seventeen in Aarau. Afterward, there is a very short account of Einstein’s life in Aarau and following this, we find Einstein’s famous thought experiment of chasing a light beam: “If a person could run after a light wave with the same speed as light, you would have a wave arrangement which could be completely independent of time. Of course, such a thing is impossible.” The authors tell the reader that the photograph is “a formal studio portrait in the carte-de-visite style, printed on card stock,... designed for presentation. Einstein gave it to his lifelong friend Albert Karr-Karusi and inscribed the back (in German)” \cite{Berger}, p. 17. This text layout, repeated throughout the book (a few sentences on Einstein’s life, on his science, and with details about the photographer and the circumstances surrounding the photo), is based on DiRuggiero’s descriptions in his “Einstein Archives” website, on quotations from Walter Isaacson’s biography of Einstein for the general reader, \emph{Einstein: His Life and Universe} \cite{Isaacson}, and on a few other books and articles, making the book reminiscent of an elegant exhibition catalog.

When looking at the book, it is as if the photographs and the handwritten documents are the actual original documents and photographs of Einstein in true size. The authors’ reproduction of original photographs of Einstein in true size speaks to the book’s great archival merits, and the book may interest collectors and dealers. Archivists are interested in such topics and questions as the photographer who “took this rare and dramatic photograph of Einstein against a dark background” and that Einstein “had difficulty writing on the photo’s glossy surface” \cite{Berger}, p. 156. Or was the photo printed on card stock? What does the inscription on the photo read? What is the difference between the styles of photos of Einstein taken by the different photographers mentioned in the book, including Trude Fleischmann, Lotte Jacobi, Philippe Halsman, Yousuf Karsh, and Alan Windsor Richards? Were there changes in the autographs made by Einstein and other people? These questions are addressed in the book.

Berger and DiRuggiero explain the letters and manuscripts and simplify the text, making the book readable and understandable for a general audience. They have explained elsewhere that a reader does not have to have a scientific background to pick the book up and enjoy it and be inspired by Einstein as much as he has inspired them \cite{Berger2}. But the book also gives us a glimpse into the world of collectors and dealers of Einstein’s items, because the authors use a language and a style that are unfamiliar both to historians and philosophers of science and to the general reader.

That said, from the scholarly perspective of the historian and philosopher of science or the physicist, the photographic accuracy and archival details come at the expense of the quality of the historical, philosophical, and scientific interpretation of the original documents. 

Hanoch Gutfreund begins his foreword to Berger and DiRuggiero’s book by saying: “\emph{Einstein: The Man and His Mind} is a unique and valuable addition to the avalanche of books that have been published on different aspects of the Einstein phenomenon” \cite{Berger}, p. 12. A central element of the so-called "Einstein phenomenon" is the Einstein myth, that is, that Einstein’s theory of relativity is incomprehensible. Berger and DiRuggiero proliferate this myth. They write, “the difficulty in understanding relativity had become the subject of humor. In an apocryphal story, when Sir Arthur Eddington was asked how it felt to be one of only three people in the world to comprehend relativity, he paused in responding, leading his interviewer to ask, ‘What is wrong, Mr. Eddington?’ Eddington replied: ‘I’m sorry, I was just wondering who the third person is’” \cite{Berger}, p. 50. However, this is, first, not an apocryphal story and, second, not an accurate depiction of what Eddington said.

Eddington’s humorous comments were made just a day before Einstein became a world celebrity. On November 6, 1919, a special joint meeting of the Royal Astronomical Society and the Royal Society of London was convened in London to report on the results of the British Expedition confirming Einstein’s prediction concerning the deflection of light by a gravitational field. The results were within the limits of experimental errors of Einstein’s value of 1.7 seconds of arc but were beyond the limit of experimental errors of the Newtonian value of 0.85 seconds of arc. Still, physicist Ludwig Silberstein was skeptical; he said that only after a careful study of Eddington’s plates would he be prepared to say that they confirm the prediction that light is deflected in accordance with Einstein’s law of gravitation. Eddington recalled that as the meeting was dispersing, Silberstein came up to him and said, “Professor Eddington, you must be one of the three persons in the world who understand general relativity.” Eddington objected to this statement. When Silberstein responded, “Don’t be modest, Eddington,” Eddington (being cynical toward Silberstein) replied, “On the contrary, I am trying to think who the third person is!”\cite{Chandrasekhar}, p. 20.

Furthermore, in my opinion, this book is an exaggerated genuflection to Einstein as a male genius, as often happens with an entire book centered on a male celebrity. He is described as “the handsome, smartly attired Einstein in a thoughtful mood,” “a fashionably dressed and stately Einstein,” “a handsome, deeply tanned, fifty-eight-year-old Einstein,” and “a handsome and serious Einstein bathed in sunshine” \cite{Berger}, pp. 62, 122. Einstein’s wife Elsa reportedly responded to being shown sophisticated equipment used to explore the shape and behavior of the universe by saying, “Well, my husband does that on the back of an envelope,” and elsewhere, asked whether she understood relativity, she answered without hesitation, ‘No, but I understand Professor Einstein’” \cite{Berger}, p. 93. These all feed into the stereotype of the male character as handsome and knowledgeable, while in a misogynistic extension of the “Einstein Phenomenon,” women cannot comprehend the genius male’s complex physics.

As a side effect of the hero worship of Einstein, there is now a booming market in Einstein’s letters and documents. Einstein’s autograph letters and holograph manuscripts bring top prices at world auctions. Einstein’s famous “God Letter” is worth more than a medieval document that has escaped the ravages of centuries. The high price is not due to rarity; Einstein penned many handwritten letters, and people tended to save them, for obvious reasons. Thus, Einstein’s handwriting is not superlatively rare. But handwritten notes penned by Einstein have been sold at auction for hundreds of times their presale estimated value.

Although these matters are the concern of the autograph and holograph dealers and collectors, there is a vicious circle here because, as Gutfreund writes in his foreword, “The authors’ gracious gesture to contribute the royalties from this book to the Albert Einstein Archives at the Hebrew University of Jerusalem is in line with their decision to dedicate their book to Einstein’s memory” \cite{Berger}, p. 14. Berger and DiRuggiero have made their book in conjunction with the Albert Einstein Archives. With all the good intent in the world, if collectors and booksellers are making books on Einstein in conjunction with the Einstein Archives and are donating royalties from these books to the Einstein Archives, the Einstein Archives then implicitly participate in the collectors’ market of Einstein’s letters and documents, thus legitimating the fact that many Einstein documents are in private hands.

The relationship between archives and collectors is quite subtle. Curators and archivists are not passive; they are always active and alert and continually seek new original manuscripts, constantly prepared to acquire hitherto unknown original documents even (or especially) from collectors. The point is that original documents do not endure forever. They are subject to the attrition of time. Subtle changes in temperature and light, pollution, dust, mold, and humidity cause damage to documents. To protect documents from long-term deterioration, they need to be stored in special conditions. They further need special care and attention, because items must be repaired and treated to improve their condition. Institutions have the facilities, technology, and expertise to preserve original manuscripts. Collectors and booksellers most likely cannot provide these conditions. I therefore believe that if we wish to “dedicate” something “to Einstein’s memory,” then Einstein’s original letters and documents should not be in private hands.

Einstein asked to be cremated so people would not come to worship at his grave. Instead, authors have flooded the market with books worshiping Einstein. A recent proliferation of books try to present to the reader original letters, notebooks, and manuscripts of Einstein and to add interesting stories and anecdotes that tie them to Einstein’s scientific work. The trouble with the ongoing success of the Einstein industry is that now collectors and dealers of Einstein’s holographs are going to publish elegant catalogs of Einstein’s items under the auspices of the Albert Einstein Archives. The bare truth is that very few people besides the most dedicated Einstein collectors and archivists can truly benefit from original photographs inscribed by Einstein. Neither can a reproduction of autographs and holographs by Einstein be of much use for scholars.

An example can sum up both problems—of hero worship and scholarly utility. On pages 69-71, Berger and DiRuggiero provide a true-to-size reproduction of a holograph letter sent by Einstein to his closest friend, Michele Besso, on Einstein’s unified field theory of 1929. We see the entire letter and the envelope with Einstein’s writing on it. Berger and DiRuggiero wax eloquent about the sensation caused by the publication of Einstein’s “most ambitious work” \cite{Berger}, p. 68. Yet scholars recognize that Einstein resorted to wishful thinking here because the majority of the scientific community was doing research in quantum mechanics and nuclear physics. Indeed, Einstein’s unified field theory was riddled with problems. In addition, the full, unabridged contents of the letter can readily be found as letter 93 in Pierre Speziali’s 1972 book, \emph{Albert Einstein and Michele Besso Correspondence 1903-1955}. \cite{Besso}, and in the Collected Papers of Albert Einstein \cite{CPAE}, doc. 358. So, as a historian and philosopher of science, why should it matter in any way that I read the letters and equations written in Einstein’s handwriting? For Einstein scholars like me, it is enough to read the documents in one of the above, readily available sources.

My impression is, therefore, that a book on Einstein showing reproductions of so-called never-before-seen autographs and holograph items is more like a unique exhibition catalog for collectors of Einstein’s items and archivists than a useful book on Einstein for scholars.

\section*{Acknowledgement}

\noindent This work is supported by ERC advanced grant number 834735.

\end{document}